\documentclass[fleqn,10pt]{wlscirep}
\usepackage[utf8]{inputenc}
\usepackage[T1]{fontenc}
\usepackage{nameref}
\title{Inferring Skin-Brain-Skin Connections from Infodemiology Data using
  Dynamic Bayesian Networks}

\author[1,*]{Marco Scutari}
\author[2, 3]{Delphine Kerob}
\author[2]{Samir Salah}
\affil[1] {Istituto Dalle Molle di Studi sull'Intelligenza Artificiale (IDSIA),
  Lugano, Switzerland}
\affil[2]{La Roche-Posay Dermatological Laboratories, Levallois-Perret, France}
\affil[3]{Department of Dermatology, AP-HP Saint-Louis Hospital, Paris, France}

\affil[*]{scutari@bnlearn.com}

\begin{abstract}

  \paragraph{Background:} The relationship between skin diseases and mental
  illnesses has been extensively studied using cross-sectional epidemiological
  data. Typically, such data can only measure association (rather than
  causation) and include only a subset of the diseases we may be interested in.

  \paragraph{Objective:}  In this paper, we complement the evidence from such
  analyses by learning an overarching causal network model over twelve health
  conditions from the Google Search Trends Symptoms public data set.

  \paragraph{Methods:} We learned the causal network model using a dynamic
  Bayesian network, which can represent both cyclic and acyclic causal
  relationships, is easy to interpret and accounts for the spatio-temporal
  trends in the data in a probabilistically rigorous way.

  \paragraph{Results:} The causal network confirms a large number of cyclic
  relationships between the selected health conditions and the interplay between
  skin and mental diseases. For acne, we observe a cyclic relationship with
  anxiety and attention deficit hyperactivity disorder (ADHD) and an indirect
  relationship with depression through sleep disorders. For dermatitis, we
  observe directed links to anxiety, depression and sleep disorders and a cyclic
  relationship with ADHD. We also observe a link between dermatitis and ADHD and
  a cyclic relationship between acne and ADHD. Furthermore, the network includes
  several direct connections between sleep disorders and other health
  conditions, highlighting the impact of the former on the overall health and
  well-being of the patient. The average $R^2$ for a condition given the values
  of all conditions in the previous week is 0.67: in particular, 0.42 for acne,
  0.85 for asthma, 0.58 for ADHD, 0.87 for burn, 0.76 for erectile dysfunction,
  0.88 for scars, 0.57 for alcohol disorders, 0.57 for anxiety, 0.53 for
  depression, 0.74 for dermatitis, 0.60 for sleep disorders and 0.66 for
  obesity.

  \paragraph{Conclusions:} Mapping disease interplay, indirect relationships,
  and the key role of mediators, such as sleep disorders, will allow healthcare
  professionals to address disease management holistically and more effectively.
  Even if we consider all skin and mental diseases jointly, each disease
  subnetwork is unique, allowing for more targeted interventions.

\end{abstract}
\begin{document}

\flushbottom
\maketitle

\thispagestyle{empty}

\section*{Introduction}

Skin diseases and mental illnesses have been extensively studied. However, they
are commonly investigated in isolation: the interplay between different skin
diseases, between mental illnesses, and between skin diseases and mental
illnesses are ignored, limiting our understanding of their aetiology. Skin and
brain may interact in four ways: skin-to-skin, brain-to-brain, skin-to-brain and
brain-to-skin. Skin-to-skin interactions, or skin disease associations, may
arise because of the general altered skin barrier function shared by all
inflammatory skin diseases\cite{lee06}; because of the use of topical drugs that
induce an alteration of the skin barrier function or other skin
reactions\cite{zhou22,rocha18}; and because of common aggravating mental and
environmental risk factors\cite{evans20,roberts21}.

As for brain-to-brain interactions, there is a growing consensus among
psychiatrists that the boundaries between mental disorders, which often overlap
in signs and symptoms, are not clear: a recent study shows that mental diseases
share a large number of genetic variants\cite{marshall20}. Anxiety and
depression, for example, have a genetic correlation of 0.79, while ADHD has a
correlation of 0.39 with anxiety and 0.52 with depression. Clinical practice
should take into account these overlapping genetic contributions to reduce
diagnostic errors and treatment effect heterogeneity on psychiatric disorders.

Skin-to-brain interactions have largely been investigated through the evaluation
of the impact of skin diseases on mental health, mainly anxiety, depression and
attention deficit hyperactivity
disorder\cite{uhlenhake10,purvis06,samuels20,moller18,patel19,yaghmaie13,chen14}.
This relationship has also been studied indirectly through the impact of skin
diseases on symptoms directly related to mental illnesses such as depressive
symptoms, social isolation and loneliness\cite{barankin02,hong08,yew20}. A
second important pathway is through mediators like the quality of sleep in
connection with pruritus\cite{lavery16,hawro20,chamlin05,dahl95}.
Over-representation of sleep disorders has been observed in patients with
psoriasis, atopic dermatitis, hidradenitis suppurativa and
vitiligo\cite{mouzas08,kaaz18,gupta16,chang18}. More recent works study the
reverse effect of sleep deprivation on skin disease through the bi-directional
relationship between sleep and the immune system\cite{myers21}. This link is
thought to contribute to the chronic inflammation observed in many skin
diseases. Therefore, dermatologists should emphasise sleep hygiene in their
practice.

Dermatologists studying brain-to-skin interactions have been arguing that
stress, anxiety and depression can aggravate or precipitate the onset of most
inflammatory skin diseases. The COVID-19 pandemic provided strong evidence
supporting this hypothesis, with dermatologists reporting an increase in the
incidence of flares during this period\cite{shah20}. Unfortunately, this effect
is difficult to quantify because patient access to medical treatment was
restricted during lockdown periods when stress, anxiety and depression were most
likely to develop. The hypothalamic pituitary adrenal (HPA) axis is responsible
for responding to psychological stress, which produces both pro- and
anti-inflammatory effects on the skin\cite{evans20} in turn. Initially, the
release of pro-inflammatory cytokines by the corticotropin-releasing hormone
(CRH) starts a quick inflammatory process. However, CRH also triggers a slower
anti-inflammatory process that leads to the release of glucocorticoids
(cortisone, cortisol). Studies have shown that pro-inflammatory cytokines induce
mast cell activation, promoting immune dysregulation and neurogenic
inflammation\cite{snast18,arck06}. They are known to play a role in allergic
reactions as well\cite{galli10}. Inflammatory skin diseases share a common link
with the quality of the skin barrier function: the enzyme
11$\beta$-hydroxysteroid dehydrogenase type I, which is responsible for the
transformation of cortisone (inactive form) into cortisol (active form), is a
marker of barrier function impairment\cite{choe18}. Anti-psychological stress
interventions like SSRI reduce both enzyme concentration and improved barrier
function.

Psychological stress is not only an aggravating factor for skin diseases but may
also trigger diseases like vitiligo\cite{papa98}, psoriasis\cite{ott09},
seborrheic dermatitis\cite{misery07}, trichotillomania, excoriation disorder and
delusions of parasitosis\cite{rahman22} in individuals with genetic
susceptibility.

Complementing this epidemiological evidence with a model of the interplay
between these four classes of interactions is crucial to improving the diagnosis
and treatment of skin diseases and mental illnesses in medical practices.
Modelling this interplay using epidemiological data is very difficult because of
the lack of comorbidity studies with a longitudinal design. In this paper, we
consider the search trends infodemiologic data available in the Google COVID-19
Public Data, an analysis-ready large longitudinal data set\cite{data}. It is not
the first time that this data set has been used to complement epidemiological
insights for achieving a larger sample size or for disease early
detection\cite{ginsberg09}. The intuition behind its use is that many patients
perform online searches about their putative conditions before visiting a
physician\cite{ginsberg09,riel17}. We can then assume a non-negligible
association between the frequency of online searches for specific diseases and
the actual incidence of those diseases in physicians' diagnoses. Restricting
ourselves to searches performed on Google is not a significant limitation
because of the prevalence of its use: 8.5 billion queries are processed by
Google every single day\cite{oberlo}. At the time of this writing, a search of
PubMed titles and abstracts with the keywords ``search trend'' and ``Google
search'' yields 1489 results in 2020, 59\% more than in 2019. This volume has
been maintained in 2021 and 2022. The COVID-19 pandemic has highlighted how such
data can be used to track pandemics and, more interestingly, to study
longitudinal patterns of disease progression in several healthcare
domains\cite{lampos21,lu21}. However, poor documentation practices and high
heterogeneity in methods have led to conflicting findings\cite{nuti14}. Previous
infodemiologic studies based on Google search trends relied on keyword matching,
which is likely to miss many classes of queries, and many of them did not
incorporate query translation, which may result in significant loss of
information in countries like the US where multiple languages are in common
use\cite{cerve17}. To overcome these two issues, Google leveraged the latest
advancements in natural language processing (NLP) for query classification and
translation, including state-of-the-art transformers deep neural
networks\cite{vas17}, to create the COVID-19 Public Data set. Training on large
corpora makes them more accurate in classifying diseases even with little to no
fine-tuning (respectively, zero-shot classification without additional
supervised learning\cite{yin19} and with few-shot classification with a small
number of examples\cite{ye20}).

We will use the COVID-19 Search Trends Symptoms data set, which is part of
Google's COVID-19 public data, to model the interplay between skin diseases and
mental illnesses with a causal model built on dynamic Bayesian networks (dynamic
BNs)\cite{crc21}. This class of graphical models has a unique set of advantages:
they provide a graphical representation of the interactions between diseases;
they can be learned automatically from data, from expert knowledge, or a
combination of both; and they can be used as diagnostic or prognostic support
systems because they can easily evaluate any scenario of interest. Most
importantly, unlike the more common correlation networks, dynamic BNs represent
interactions as directed arcs that can disambiguate causes and
effects\cite{bressler11,causality}. This ability is crucial in finding
appropriate targets for treatment, thus avoiding symptomatic treatment and
improving clinical outcomes, and in differentiating unidirectional cause-effect
relationships from feedback loops. As a result, dynamic BNs provide a clearer
picture of the interplay between the skin and the brain and, at the same time,
they are better suited to design treatment regimes.

\section*{Results}

\begin{figure}[t]
  \centering
  \includegraphics[width=0.65\linewidth]{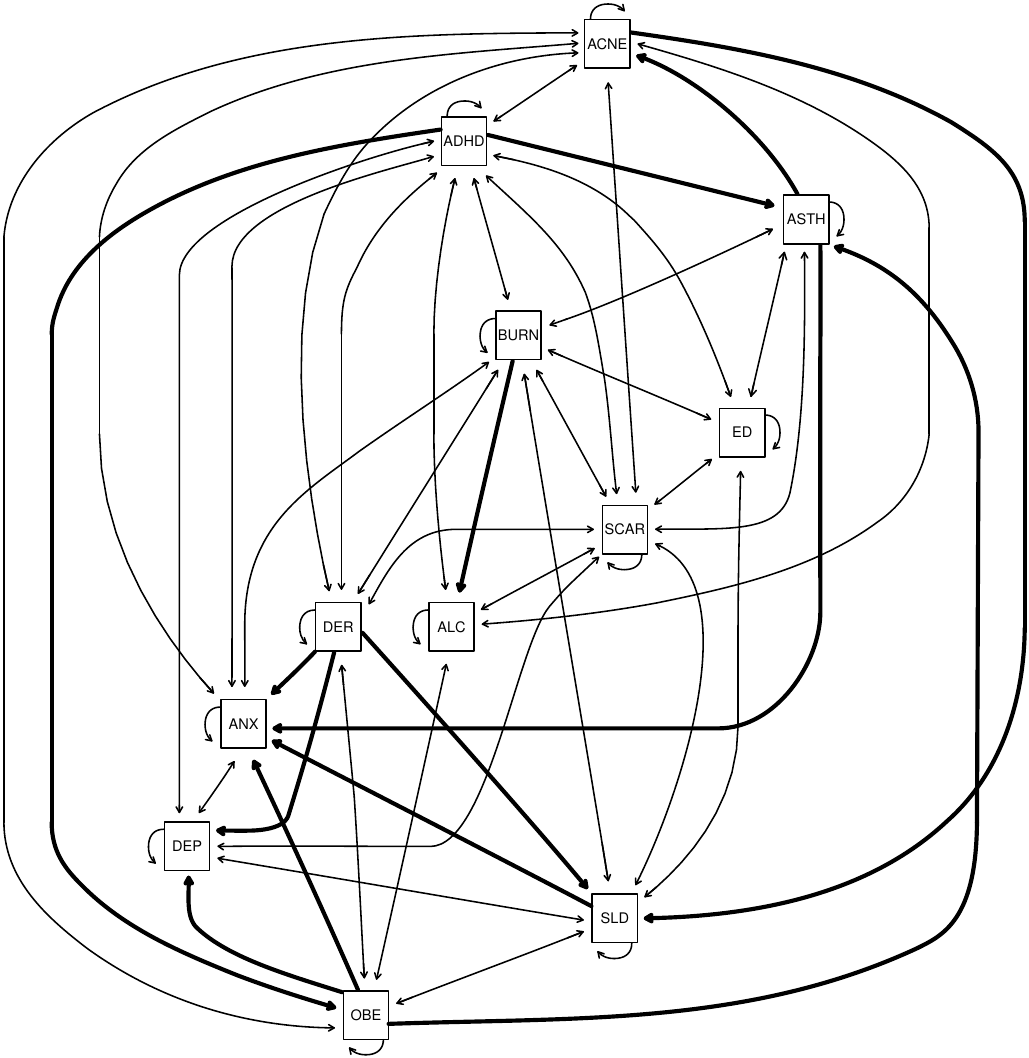}
  \caption{The dynamic Bayesian network linking skin diseases and mental
    illnesses learned from the Google COVID-19 Search Trends Symptoms public
    data set. Bidirectional arcs represent cyclic relationships, that is,
    feedback loops. Unidirectional arcs between different nodes represent
    one-way causal effects and are drawn with thicker lines for emphasis. Those
    from a node to itself represent autocorrelation through time.}
  \label{fig:dbn}
\end{figure}

The data set used in this work is publicly available at the URL listed in the
Data Availability. We consider the web search queries collected weekly from
users in the United States between 2020-03-02 and 2022-01-24 (100 weeks). We
focus on the following 12 conditions (with abbreviations): obesity (``OBE''),
acne (``ACNE''), alcoholism (``ALC''), anxiety (``ANX''), asthma (``ASTH''),
attention deficit hyperactivity disorder (``ADHD''), burn (``BURN''), depression
(``DEP''), dermatitis (``DER''), erectile dysfunction (``ED''), sleep disorder
(``SLD'') and scar (``SCAR''). We detail the mapping between these conditions
and the variables in the search trends in \nameref{spt:map}. To avoid a very
sparse data set, we remove symptoms with more than 30\% missing data. The focus
is on skin diseases, excluding very general symptoms like ``Lesions'', ``Skin
conditions'', ``Infection'', ``Inflammation'' and ``Skin ulcer'' or with
potential confusion like ``Xeroderma''. The second set of symptoms are mental
conditions with a focus on anxiety, depression and ADHD, the most studied mental
disease in dermatology. We retain asthma for its documented link with atopic
dermatitis and anxiety\cite{yaneva21,katon04}. Obesity is a comorbidity that
plays an important role in mental disorders like anxiety\cite{gariepy10}, so its
presence is justified to control confounding in some relationships. Obesity may
also trigger some skin diseases like atopic dermatitis and
psoriasis\cite{ali18,jensen16}. Erectile dysfunction is selected because it is
considered a probable consequence of mental disorders and sleep
deprivation\cite{velu22,cho19}. Considering additional health conditions might
allow the network to recover more information but is likely to produce dense
networks that are difficult to interpret and make computations cumbersome. For
each of these conditions, we retrieved the relative frequency in web search
queries of the relevant search terms from the Search Trends Symptoms data set in
Google's COVID-19 Public Data. Frequencies are measured in each county of each
US state over the period. Therefore, the resulting data set we used for the
analysis is a weekly multivariate time series over 12 health conditions for each
county and state with the relative frequencies of the web search queries
normalised at the county level. Missing data were imputed, and we confirmed the
accuracy of the imputation process to be satisfactory. (See the Methods for
details). The overall number of observations for each condition given by 2879
counties over 50 states and 100 weeks is 287900.

The dynamic BN model we learned from these data to investigate the
skin-brain-skin causal connections among the conditions above is shown in
Figure~\ref{fig:dbn}. The model should be read as follows: each node corresponds
to one of the conditions above, and arcs represent direct probabilistic
associations. Bidirectional arcs represent feedback loops, while unidirectional
arcs between different nodes represent one-way effects. Finally, arcs from a
node to itself represent each node's autocorrelation through time. Nodes not
connected by an arc are indirectly associated if we can find a path that
connects them without passing through any node corresponding to conditions we
may be controlling for, or conditionally independent otherwise. The longitudinal
data we learned the dynamic BN from allow us to give an additional causal
interpretation to the results within Granger's and Pearl's causality frameworks,
as we will discuss in more detail in the Methods.

The interactions between these conditions were learned completely from the data
to validate known relationships from the literature and discover new putative
ones. To reduce false positives (which implies including spurious arcs in the
dynamic BN in this context), we structured the dynamic BN to account for the
temporal and spatial dependence structure of the data, and we integrated model
averaging via bootstrap resampling in the learning process. Furthermore, we
tuned the learning process by penalising the inclusion of arcs in the dynamic BN
to find the optimal balance between predictive accuracy and the need for a
sparse network. The large sample size of Google's Search Trends Symptoms data
set gives us enough statistical power to detect even marginal effects. However,
an overly dense network model would hardly be interpretable from either a
qualitative or quantitative point of view: it would be difficult to examine
visually, and it would fail to formally identify which pairs of conditions are
independent or conditionally independent. Therefore, we chose the final dynamic
BN shown in Figure~\ref{fig:dbn} to have about three arcs for each condition.
Even so, the average predictive accuracy across all conditions is 99.96\% of
that of the dynamic BN learned in the same way but with no penalisation. In
absolute terms, the $R^2$ for each condition given the values of all conditions
in the previous week are 0.42 for ACNE, 0.85 for ASTH, 0.58 for ADHD, 0.87 for
BURN, 0.76 for ED, 0.88 for SCAR, 0.57 for ALC, 0.57 for ANX, 0.53 for DEP, 0.74
for DER, 0.60 for SLD and 0.66 for OBE (average 0.67). This level of predictive
accuracy suggests that the dynamic BN in Figure~\ref{fig:dbn} may be a realistic
causal model of this set of conditions.

\begin{figure}[t]
  \centering
  \includegraphics[width=0.85\linewidth]{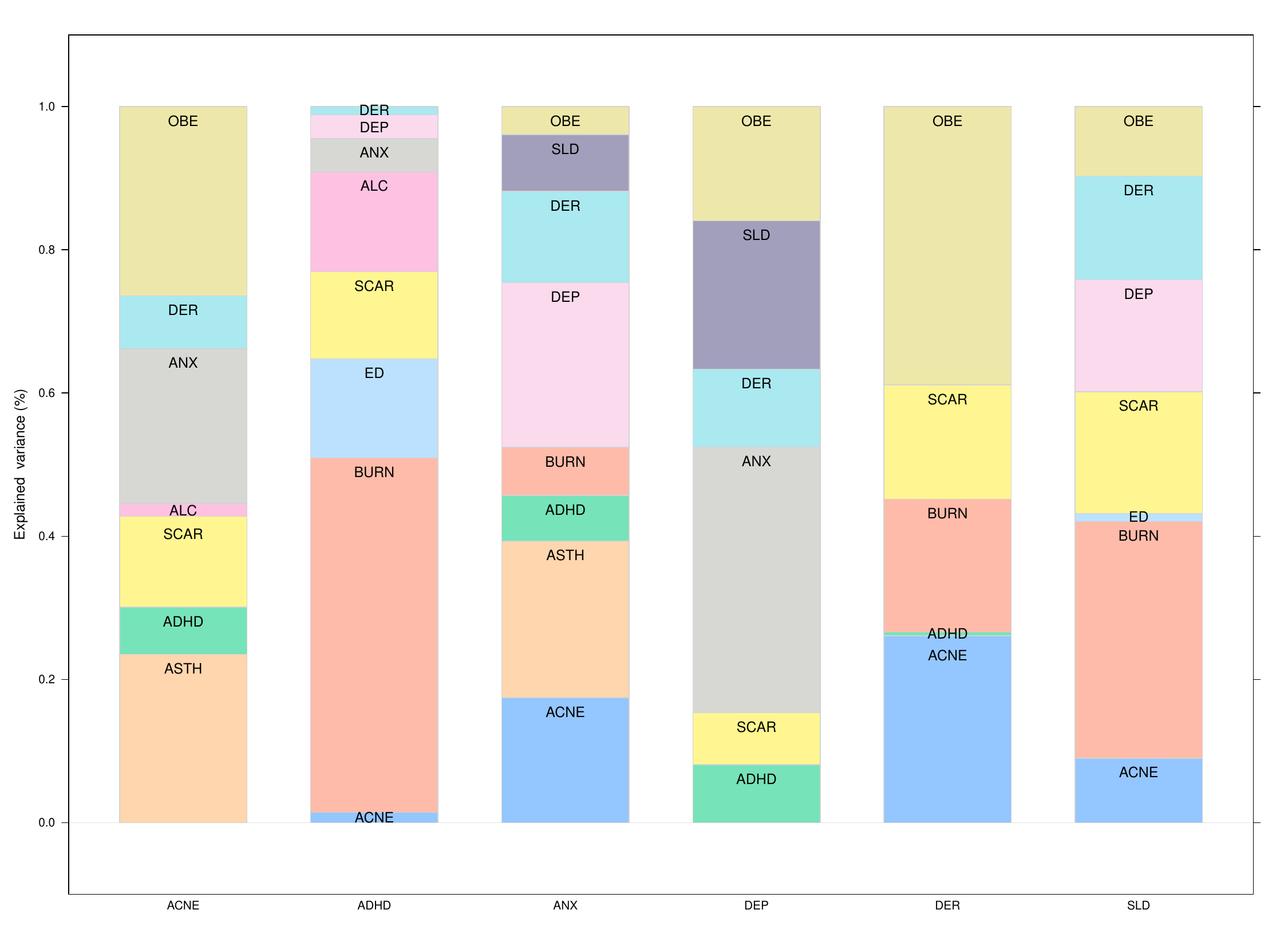}
  \caption{Proportions of the variance of some conditions of interest explained
    by their parents in the network shown in Figure~\ref{fig:dbn}, normalised by
    the total explained variance. The proportions are computed excluding the
    contribution of the condition itself from the previous time point in the
    data.}
  \label{fig:varprop}
\end{figure}

The strengths of the arcs shown in Figure~\ref{fig:dbn} are reported in
\nameref{spt:strength}. They can be interpreted as our confidence that the arcs
are statistically significant or as their probability of inclusion in the
dynamic BN. As expected, most are close to 1 (where 0 represents a complete lack
of confidence, and 1 is the strongest possible confidence). In the case of
feedback loops, we observe that both arc directions have approximately the same
strength with only two exceptions: acne and dermatitis (1 for ACNE $\to$ DER,
0.906 for DER $\to$ ACNE), and ADHD and dermatitis (0.688 for ADHD $\to$ DER,
0.996 for DER $\to$ ADHD). The difference in arc strength was 0.026 or less for
all other feedback loops. Note that the model allows us to give a causal
interpretation to the arcs within the frameworks of Granger's and Pearl's
causality, so we can interpret these arc strengths as the probability of causal
effect between the variables they connect.

We show in Figure~\ref{fig:varprop} the proportions of variability of ACNE, ANX,
DER, DEP, SLD and ADHD explained by the respective parents, normalised by the
total variance explained for the condition: these conditions are of particular
interest to us. (The raw proportions are shown in \nameref{spf:explained}.) The
proportion of variability explained by each parent measures the magnitude of its
direct causal effect on the trait and complements the corresponding arc strength
(which only measures how confident we are that a causal effect exists at all).
This information is especially useful in the case of the dynamic BN from
Figure~\ref{fig:dbn} because all the arc strengths are equal to or close to 1,
regardless of the associated effect sizes, because of the large sample size of
Google's Search Trends symptoms data set.

In particular, we note that ASTH accounts for a nontrivial share (23.5\%) of the
explained variability of ACNE in the following week, less than OBE (26.4\%) but
more than ANX (21.6\%). However, the association between the ASTH and ACNE is
not well established in the literature. We attribute this finding to the fact
that ANX is connected to all of ASTH, ACNE and OBE, which may introduce
confounding in models that do not consider them simultaneously. For instance,
such confounding may arise because of the feedback loop between ACNE and OBE
(which are both connected to ANX) unless OBE is controlled for. Our model
supports the hypothesis that ASTH and ACNE are only weakly associated (in the
same week) after controlling for ANX (in the previous week): if we perform
inference on the dynamic BN as described in the Methods, ASTH only accounts for
0.41\% of the explained variance of ACNE. Stratifying by ANX (low: bottom
quartile of search query frequencies; high: top quartile; average: second and
third quartiles) reveals that the share of explained variance is even smaller
when ANX is average or low (0.46\%, 0.36\%, respectively) compared to when ANX
is high (0.53\%).

Furthermore, the model supports the hypothesis that the feedback loop between
ACNE and OBE is driven by ANX: ACNE account for a sizeable share of the
explained variability of OBE (62.2\%) and vice versa (26.4\%). However, after
controlling and stratifying for ANX using BN inference, OBE only accounts for a
trivial share of the explained variability of ACNE in the same week (1.6\% for
high ANX, 1.7\% for low ANX and 1.3\% for average ANX). The same is true for
ACNE, which accounts for similar share of the explained variance of OBE.

These results confirm the cyclic relationships between skin diseases and mental
illnesses. Figure~\ref{fig:varprop} highlights the importance of each trigger
for the six health conditions. Depression is mainly driven by mental disorders
and sleep disorders, but dermatitis explains a significant proportion of its
search popularity. Anxiety triggers are more diverse: skin conditions like acne
and dermatitis play an important role. Triggers of ADHD, learned by the dynamic
BN without any reference to clinical experts or the literature, give a new
insight into the disease. It is already known that ADHD is associated with a
risk increase of burn injuries and scarring (from burns): these data suggest
that an increase in burn injuries may lead to ADHD diagnosis.

For acne, we observed a strong, direct cyclic relationship with anxiety and ADHD
and an indirect relationship with depression through sleep disorders. For
dermatitis, we observed directed links to anxiety, depression and sleep
disorders and a cyclic relationship with ADHD. We also observed a link between
dermatitis and ADHD, and a cyclic relationship between acne and ADHD. Although
not exclusive, the role of mediators like sleep disorders is confirmed in the
network with a significant contribution to anxiety and depression. Acne, burns,
scars and dermatitis directly impact sleep disorders. The learned BN visualises
multiple disease relationships in a single picture.

\section*{Discussion}

Our results confirm the interplay between skin diseases and mental illnesses at
the infodemiological level. Several skin-to-skin, brain-to-brain, skin-to-brain
and brain-to-skin relationships are highlighted in the model. It is interesting
to see well-known clinical relationships reproduced in the dynamic BN and put
into a larger context with deeper explanations. The data we used for the study
uniquely allow for these results by providing almost complete coverage of the US
population, with high-quality labelling of web searches, and for an extended
period of time.

At the same time, our analysis has two main limitations. Firstly, the
correspondence between infodemiological and epidemiological information is a
crucial assumption that is difficult to test in practice but that appears to be
supported by the other studies using the Search Trends Symptoms
data\cite{info1,info2,info3}. Even so, we cannot draw clinical conclusions about
single individuals because the causal relationships we observe at the level of
the aggregate data may not necessarily hold for individuals \cite{morgenstern}.
Secondly, there may be latent confounders that introduce spurious arcs or causal
directions in the dynamic BN. The almost complete coverage of the web searches
of the US population suggests that neither sampling bias nor systematic patterns
of missing values affect our causal analysis. Likewise, latent variables that
act as confounding factors at the level of individuals are not necessarily
confounding factors at the level of the aggregate data we are analysing.
However, there may be conditions that we have not included in the analysis that
introduce confounding at the level of the aggregate data. There are also
patterns of confounding specific to aggregate data we should be aware
of\cite{greenland}.

A directed link between two conditions in a dynamic BN can represent different
types of associations. It may indicate direct causation, where one condition
directly influences the other. Alternatively, the link could reflect a temporal
dependency without causation, where the two conditions are closely associated in
time due to shared genetic factors or similar environmental exposures. Such
temporal associations might lead to the sequential manifestation of symptoms,
which could be mistakenly perceived as a causal link. Additionally, the observed
relationship could be the result of an unobserved confounding factor,
introducing bias into the interpretation. It is important to recognise that one
condition may not cause the other to occur; rather, the link could suggest that
one condition may activate or exacerbate symptoms of an already existing
condition.

To illustrate the types of associations, we can consider the arc from ADHD to
asthma. This link could imply direct causation if ADHD-related behaviours, such
as hyperactivity, or treatments aggravate pre-existing asthma
symptoms\cite{tsai14,gong14}. On the other hand, the link might reflect a
temporal association without direct causation, where ADHD and asthma appear
together more frequently due to shared genetic markers\cite{zhu19} or
environmental factors\cite{fluegge18} without one condition causing the other,
although ADHD symptoms may manifest first. Additionally, an unobserved factor
like socio-economic status could confound the relationship, suggesting a link
where none exists\cite{russell15,busby21}.

Nevertheless, the large number of feedback loops supports the existence of
vicious circles in which diseases exacerbate each other until treated
appropriately. The dynamic BN cannot elucidate the starting point of these
circles but emphasises the need for more holistic disease management for
dermatologists and psychiatrists. Dermatologists should take into account the
mental health of their patients, and psychiatrists should take into account the
skin problems of their patients.

The results also highlight the essential role of key mediators, like sleep
disorders, that establish a bridge between the skin and the brain. We should not
ignore these disorders if we want to act effectively on skin and mental health.
Furthermore, not controlling for comorbidities like obesity may lead to spurious
conclusions, hiding the existing relationships.

In addition, the dynamic BN has confirmed the genetic correlation between
depression, anxiety, and ADHD\cite{marshall20}, illustrated through
bi-directional arcs. The fact that these disorders share numerous signs and
symptoms may lead to additional spurious associations, potentially due to
diagnostic errors\cite{marshall20}. In the case of acne and dermatitis, these
inflammatory diseases are linked by their common feature of altered skin barrier
integrity\cite{lee06}, and they may be triggered by similar environmental
conditions\cite{evans20,roberts21}. Acne treatments are also recognised for
their potential to irritate the skin \cite{zhou22,rocha18}. While the direct
link between acne and sleep disorders has been less explored than that between
dermatitis and sleep disorders, some studies suggest a connection between the
severity of acne and sleep disorder\cite{schrom19}. As our data are continuous,
we can capture these kinds of associations.

A less evident bi-directional relationship is acne with scars. The acne-to-scars
direction is evident and known as severe forms of acne, such as nodules and
cysts, can lead to skin damage and subsequent scarring\cite{connolly17}. The
scars-to-acne direction is much less evident, but explainable: scar healing can
lead to itchiness and may cause irritation and rash\cite{robson01,farrukh20}.
Such skin reactions can result in the formation of papules, a situation often
seen with hypertrophic scars and keloids.

Even if we consider all skin and mental diseases jointly, each disease
subnetwork is unique, allowing for more targeted interventions. The conditional
independence property of BNs allows for this kind of focus without loss of
information.

In this work, we also wanted to raise awareness of the importance of measuring
causality with appropriate study designs and statistical methods leveraging
multiple conditions in longitudinal monitoring and allowing feedback loops to
reproduce the natural cycle of human health. This may significantly reduce the
number of measured associations and highlight a focused set of preferential
targets for intervention.

The second important objective of this study was to provide guidelines for
better use of search trends data to ensure robust findings. Firstly, query
classification using a keywords approach may fail to capture relevant
information, leading to low reproducibility across researchers. Using the latest
AI breakthroughs in natural language processing for query translation and
classification will ensure better reproducibility of studies.

Secondly, choosing models that can capture the main features of search trends
data is necessary to avoid several sources of bias. We provided a detailed
methodology to deal with missing data, space and time dependencies, the lack of
sparsity in the network due to the size of the data set, model interpretability
and other considerations.

The marked discrepancies between the conclusions of studies dealing with this
type of data can be attributed to how queries were classified and processed.
Standardising these two tasks will demonstrate the high potential of these data
to complement clinical evidence for a more positive impact on public health.

\section*{Materials and Methods}

All analyses were performed using R\cite{rcore} and the packages
nlme\cite{pinheiro00}, lme4\cite{bates15}, imputeTS\cite{moritz17} and
bnlearn\cite{jss09}.

\subsection*{The Search Trends Symptoms data}

Google released the Search Trends Symptoms data\cite{symptoms-data} in September
2020 as part of the COVID-19 Public Data sets. It includes aggregated county-
and state-level Google search frequencies for 422 symptoms and conditions that
might be related to COVID-19. This data set is unique in that, as mentioned in
the Introduction:
\begin{itemize}
  \item It was aggregated both by day and by week, and normalised by scaling the
    symptom search count with the total search activity in the county or state.
    Therefore, it does not require further preprocessing to make it homogeneous
    across time and space. Furthermore, the data is stationary because of the
    normalisation.
  \item Search terms were disambiguated and aggregated across languages and
    synonyms using Google's NLP models\cite{vas17}.
  \item Missing data in the weekly frequencies were imputed without loss of
    information from the daily frequencies wherever
    possible\cite{symptoms-data-short}.
  \item The prevalence of Google's use in the US \cite{oberlo} reduces the
    risk that its web searches are a biased sample that is not representative
    of those of the US population.
\end{itemize}
As a result, it has proven an effective tool to construct predictive models for
various conditions in infodemiology\cite{info1,info2,info3}. This is not
surprising because 95\% of Americans use the Internet\cite{pew}, and of those
$87.2\%$ use Google: the Search Trends Symptoms data set is almost a
population-level longitudinal data set. Therefore, we can consider it unaffected
by common sources of sampling bias such as age, race, gender, socio-economic
status, or geographical location.

Specifically, we considered the weekly search frequency data collected in the US
at the county level from March 2, 2020, to January 24, 2022. This subset of the
data spans 2879 counties and 100 time points (weeks) and contains no missing
data. Therefore, it takes the form of a multivariate time series over the 12
conditions we are studying (OBE, ACNE, ALC, ANX, ASTH, ADHD, BURN, DEP, DER, ED,
SLD and SCAR) for each county. These conditions are completely observed in the
chosen time period, which allows us to rule out confounding due to systematic
missingness.

\subsection*{Missing data management}

The Search Trends Symptoms data set originally contained missing data, either
single data points or whole time series for specific counties and periods. The
proportion of missing data in individual conditions ranged from 0\% to 16\%. We
explored different methods to impute them for both individual time series
(exponentially weighted moving average, interpolation) and multivariate time
series (Kalman filters) using the imputeTS R package. To assess their
performance, we used the complete observations and artificially introduced 2\%,
5\%, 10\% and 20\% missing values individually and in batches of four to
simulate both random missingness and lack of measurements over one month. All
combinations of missing data patterns, proportions, and algorithms produced
average relative imputation errors between 10\% and 14\%, suggesting that all
imputation methods perform similarly (\nameref{spf:imputation} and
\nameref{spf:imputation2}). Finally, we chose an exponentially weighted moving
average imputation because of its simplicity. Imputing some combinations of
conditions and counties was impossible because of the insufficient number of
observed values; we dropped them from the data, reducing the overall sample size
to 287866.

\subsection*{Spatio-Temporal dependence structure of the data}

The Search Trends Symptoms data set was collected over time and in different US
states and counties and, therefore, presents marked spatio-temporal patterns of
dependence between observations. We will summarise them here because they are
crucial in our modelling choices.

To be consistent with the assumptions of the dynamic Bayesian network we will
describe below, we quantify the magnitude of spatial and temporal dependencies
with the proportion of the variance of the conditions that they explain as
random effects in a mixed-effect model\cite{pinheiro00}. For spatial
dependencies, we further distinguish between the variance explained by the state
and by the county. For time, we assume an autocorrelation of order 1 (that is,
each condition is correlated with itself at the previous time point). The
proportions for each condition and the average over all conditions are shown in
\nameref{spf:variance}. On average, states explain 12\% of the variance of the
conditions (\mbox{min: 7\%}, \mbox{max: 16\%}), counties explain 49\%
(\mbox{min: 23\%}, \mbox{max: 84\%}), and counties together with autocorrelation
explain 64\% (\mbox{min: 35\%}, \mbox{max: 86\%}).

\subsection*{Dynamic Bayesian networks}

Bayesian networks (BNs)\cite{crc21} are a graphical modelling technique that can
leverage data combined with an expert's prior knowledge to learn multivariate
pathway models. The graphical part of the BN is a directed acyclic graph (DAG)
in which each node corresponds to a variable of interest (here, one of the
conditions) and in which arcs (or the lack thereof) elucidate the conditional
dependence (independence) relationships between the variables. The implication
is that each variable depends in probability on the variables that are its
parents in the DAG: as a result, the joint multivariate distribution of the data
factorises into a set of univariate distributions associated with the individual
variables. This property allows automatic and computationally efficient
inference and learning of BNs from data and has made them popular for analysing
clinical data\cite{cervical20,scher15,mcnally17}. In particular, BN inference
can automatically validate arbitrary hypotheses: in the simplest instance, a BN
is used as a generative model, and hypotheses are validated by stochastic
simulation.

To account for the spatio-temporal nature of the data, we use a particular form
of BN called a dynamic BN in which each variable appears in the DAG as a
separate node at each time point. The key advantage of dynamic BNs is that,
unlike classical BNs, they can represent feedback loops by allowing each
variable in a pair to depend on the other across time points: for instance,
$X_{1, t_0} \to X_{2, t_1}$ and $X_{2, t_0} \to X_{1, t_1}$ implies a feedback
loop between $X_1$ and $X_2$ across two consecutive time points $t_0$ and $t_1$.
Arcs are only allowed to point forward in time from a variable measured at $t_0$
and one measured at $t_1$ to ensure that they are semantically meaningful and to
be able to relate the dynamic BN to the Granger\cite{bressler11} and
Pearl\cite{causality} causality frameworks. We disallow ``instantaneous arcs''
between variables measured in the same time point for similar reasons. With
these restrictions, we can construct a directed cyclic graph that can represent
feedback loops from the DAG by folding the nodes corresponding to the same
variable at different time points into a single node. As a result, pairs of arcs
like $X_{1, t_0} \to X_{2, t_1}$ and $X_{2, t_0} \to X_{1, t_1}$ are transformed
into cycles of the form $X_1 \leftrightarrows X_2$; and arcs like $X_{1, t_0}
\to X_{1, t_1}$ become loops that express autocorrelation. In principle, we
could extend this construction to consider arcs from time points further in the
past, such as $X_{1, t_{-1}} \to X_{2, t_1}$. However, the resulting dynamic BN
would quickly lose interpretability because the number of nodes (arcs) would
increase multiplicatively (exponentially) and because we would lose the ability
to display an interpretable cyclic graph.

We should also critically examine the theoretical assumptions underlying causal
inference in BNs: the faithfulness condition and the absence of latent
confounders. The faithfulness condition requires that the observed probabilistic
dependencies are entirely due to the causal structure of the network. The lack
of latent confounders, defined as unobserved variables that are parents of at
least two observed variables, avoids the risk of edges representing spurious
causal effects. Furthermore, the data we learn the BN from should be
representative of the population we would like to study and in sufficient
quantity to ensure that we have enough statistical power to identify causal
effects. They should also be free from sampling bias and systematic patterns of
missing values: both act as latent confounders.

As for the distributions of the variables, we assume that 1) the search query
frequencies in any given week can depend in probability on those in the previous
week but not on older ones, and 2) the data are stationary, so we only need to
model the dependence between two generic consecutive times $t_0$ and $t_1$.
These assumptions allow us to parameterise the dynamic BN similarly to a vector
auto-regressive time series\cite{var05}: each condition $X_i$ in each county is
therefore modelled using a linear regression model of the form
\begin{equation*}
  X_{i, t_1} = \mu_{i, t_1} + pa(X_{i, t_1})\beta_{i, t_1} + \varepsilon_{i, t_1}
\end{equation*}
at time $t_1$ and
\begin{equation*}
  X_{i, t_0} = \mu_{i, t_1} + \varepsilon_{i, t_0}
\end{equation*}
at time $t_0$. Here $pa(\cdot)$ denotes the parents of the variable, $\mu_{i}$
is the intercept, the $\beta_{i}$ are the associated regression coefficients,
and the $\varepsilon_{i}$ are normally distributed residuals with mean zero and
some variance specific to each node. The contribution of each parent to the
linear regression can thus be naturally measured by the associated explained
variance in the model (ANOVA). Conditions have different scales arising from the
different popularity of the corresponding search queries: to make them easier
to compare in the Results, we normalise the variance explained by each parent
into a proportion (that is, we divide it by the total explained variance of all
parents). In doing so, we omit the contribution of the condition itself: in the
auto-regressive model we are considering, auto-correlations are strong enough to
make the contributions of other conditions appear less significant for purely
numerical reasons.

In addition to accounting for the time dimension, the dynamic BN incorporates
the spatial structure of the data. Assume that each condition has a different
baseline value in each county that does not change over time. Then the local
distribution of each condition at $t_1$ is regressed against itself at $t_0$
(same county, previous time point). The different baseline for the state then
appears on both sides of the equation and can be accounted for in the regression
model.

We learned the dynamic BN from the data by choosing the optimal DAG that
maximises the penalised log-likelihood
\begin{equation*}
  PL(D, w) = \log L(D) - w \log(n) p
\end{equation*}
where $p$ is the number of parameters of the dynamic BN and $n$ is the number of
observations in the data, using the hill-climbing greedy search
algorithm\cite{norvig09}. We then estimated the parameters of the dynamic BN
(the intercepts $\mu_{i}$ and the regression coefficients $\beta_{i}$) with the
chosen DAG using maximum likelihood. To ensure that the DAG is as sparse as
possible without sacrificing predictive accuracy, we chose the penalty
coefficient $w$ by learning the dynamic BN with $w = 1, 2, 4, 8, 16, 32, 64,
128$ on the first 52 weeks of data and then computing the average proportion of
variance explained over all conditions in the remaining weeks of data acting as
a validation set. For reference, $w = 1$ gives the Bayesian Information
Criterion (BIC)\cite{schwarz}. Larger values of $w$ penalise the inclusion of
arcs in the DAG by increasingly large amounts, effectively decreasing the value
of the penalised likelihood if the associated regression coefficients are small.

\begin{figure}[t]
  \centering
  \includegraphics[width=0.5\linewidth]{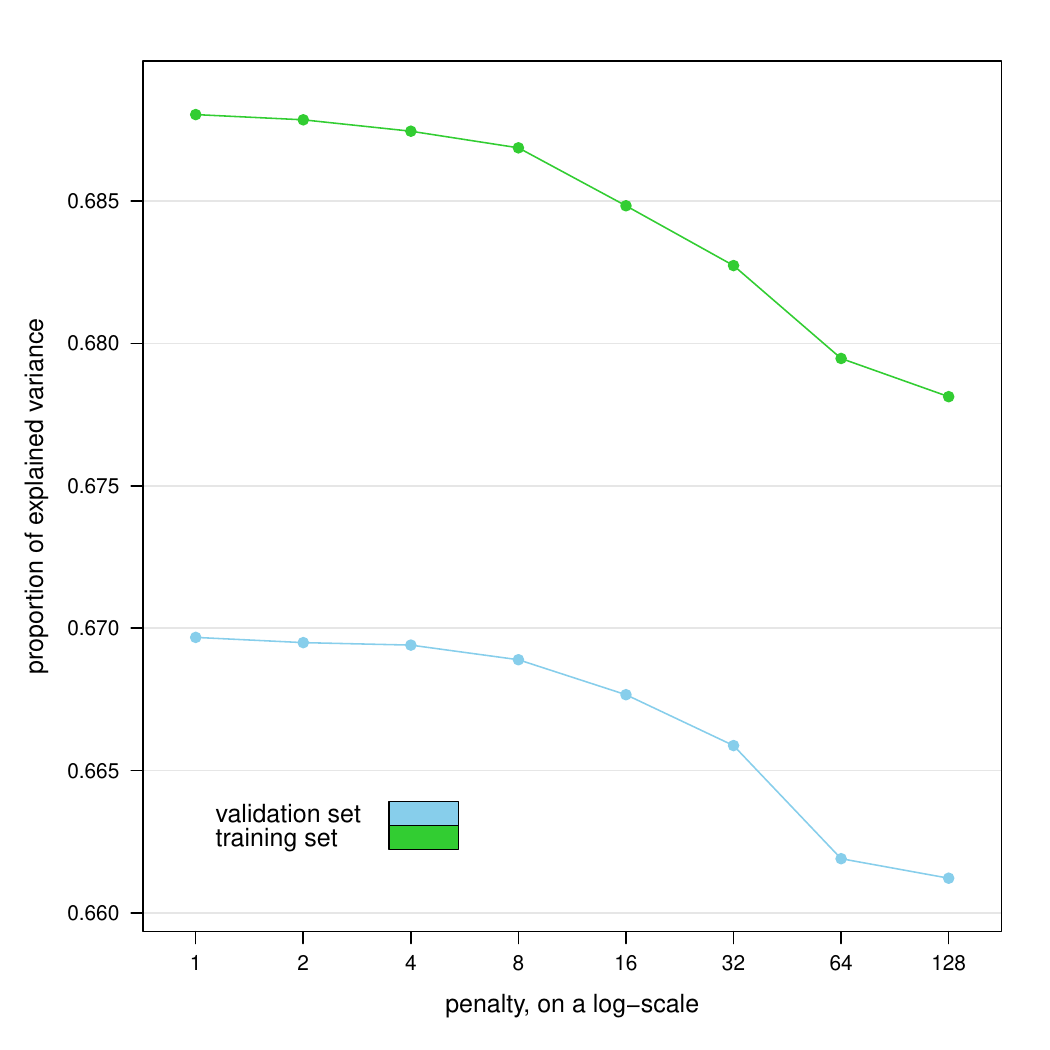}
  \caption{Average proportion of explained variance over all conditions
    explained by the dynamic Bayesian networks learned from the first year's
    worth of data (training set) in the remaining weeks (validation set) for
    various values of the penalty coefficient $w$.}
  \label{fig:selection}
\end{figure}

The resulting proportions of explained variance are plotted against $w$ in
Figure~\ref{fig:selection}. We observe no marked decrease in predictive accuracy
until $w = 4$, which we choose as the best trade-off with the sparsity of the
DAG. For reference, the dynamic BN learned with $w = 1$ has 123 arcs, while that
learned with $w = 4$ has 87 arcs with a predictive accuracy that is 99.96\% of
the former model. We also note that the variance explained by the dynamic BNs in
the validation set is not markedly different from that in the training set they
were learned from, suggesting no overfitting.

The large number of observations available on each condition gives the dynamic
BN enough statistical power to detect even marginal effects from the data and
thus to include them as arcs in the model, resulting in an overly dense DAG. The
fact that increasing $w$ and dropping many of those arcs has a limited impact on
predictive accuracy suggests that the relationships they express may be of
limited clinical relevance for diagnostic or prognostic purposes.

Furthermore, we used model averaging to reduce the potential of including
spurious arcs in the BN. We implemented it using bootstrap aggregation: we
extracted 500 bootstrap samples from the data, learned the DAG of a dynamic BN
from each of them, and then constructed a consensus DAG containing only those
arcs that appear with frequencies higher than the threshold 0.518 estimated from
the data\cite{aime11}. Each frequency measures the strength of the corresponding
arc, that is, of our confidence that the arc is supported by the data: a value
of 0 represents a complete lack of confidence, whereas a value of 1 represents
perfect confidence. Overall, the individual frequencies are estimated from the
joint arc frequencies for all possible combinations of arcs, and they are used
to compute the optimal threshold so that we do not have to manually choose its
value. If two nodes are connected with a frequency greater than the threshold,
the consensus DAG will contain an arc connecting them. The direction of that arc
will be the most common direction observed in the DAGs from the bootstrap. Arcs
are included in the consensus DAG in decreasing order of strength, and
lower-strength arcs are discarded if they introduce cycles. To further prevent
patterns in the data from impacting the consensus model, we increased the
variability of the bootstrap samples by randomising the order of the variables
and by reducing their size to 75\% of that of the original data.

Finally, we assessed the impact of the spatio-temporal structure of the data on
structure learning to motivate using dynamic BNs. For this purpose, we performed
both structure and parameter learning as described above to fit a classical
(that is, static) BN in which variables are not replicated across time points.
As a means of comparison, we learned a second static BN from the data after
removing their spatio-temporal structure with the mixed-effect model we used
above to quantify the proportion of variance explained by the counties and the
temporal autocorrelation. If we only perform parameter learning, we find that
63\% of the regression coefficients are inflated by a factor of at least two
compared to those we learn after removing the spatio-temporal structure from the
data. The sign is different for 29\% of them. If we perform structure learning
from both sets of data, we find that 71\% of the arcs differ between the two.
The reason is that a classical, static BN is a misspecified model, and the
dependence relationships between the variables are confounded with space-time
dependencies between the data points. If we take the dynamic BN in
Figure~\ref{fig:dbn} to be the correct model, the static BN learned from the raw
data has 11\% correct arcs ($X_i \to X_j$ in both networks), 50\% arcs that
should be feedback loops ($X_i \to X_j$ in the static network, $X_i
\leftrightarrows X_j$ in the dynamic BN), 8\% reversed arcs ($X_i \to X_j$ in
the static network, $X_i \leftarrow X_j$ in the dynamic BN) and 30\% spurious
arcs ($X_i \to X_j$ in the static network, no arc between $X_i$ and $X_j$ in the
dynamic BN). The proportions for the static BN learned from the data after
removing the spatio-temporal structure of the data are similar. This supports
our choice to use a dynamic BN: the large number of feedback loops in the causal
structure of the data means that any model that cannot express them will report
a large number of incorrect causal directions.

\section*{Code availability}

The code used for the analysis is publicly available at the URL:
\begin{center}
  \url{https://www.bnlearn.com/research/loreal21}
\end{center}

\section*{Data availability}

The Google COVID-19 Public Data Set is publicly available at the URL:
\begin{center}
  \url{https://console.cloud.google.com/marketplace/product/bigquery-public-datasets/covid19-public-data-program}
\end{center}
In particular, the Search Trends Symptoms data set is available at the URL:
\begin{center}
\url{https://console.cloud.google.com/marketplace/product/bigquery-public-datasets/covid19-search-trends}
\end{center}
and at the URL:
\begin{center}
\url{https://github.com/google-research/open-covid-19-data/}
\end{center}

\section*{Acknowledgements}

We thank Dr. Katrina Abuabara for her comments and suggestions on an early draft
of this paper.

\section*{Author contributions statement}

M.S. analysed the data and wrote the software for the analysis.
D.K. conceived the study and interpreted the clinical results.
S.S. conceived the study, collected the data and interpreted the clinical results.
All authors wrote and reviewed the manuscript.

\section*{Competing interests}

The authors declare no competing financial or non-financial interests.

\section*{Supplementary information}

\paragraph{Supplementary Table 1} \label{spt:map}
Mapping between the variables in the Google COVID-19 Public Data Set and the
conditions discussed in this paper. When multiple variables map to the same
condition, the search query frequencies from those variables were aggregated to
give a single overall frequency for the condition.

\paragraph{Supplementary Table 2} \label{spt:strength}
Arc strengths for the dynamic Bayesian network model shown in
Figure~\ref{fig:dbn}. ``From'' denotes the node at the tail of the arc, ``To''
denotes the node at the head of the arc, and ``Arc strength'' is the frequency of
the arcs in the bootstrapped models.

\paragraph{Supplementary Figure 1} \label{spf:explained}
Proportions of the variance of ACNE, ADHD, ANX, DEP, DER, and SLD explained by
their parents in the network shown in Figure~\ref{fig:dbn}, unnormalised. This
figure complements Figure~\ref{fig:varprop} in which the proportions are
normalised by the total explained variance for the condition.

\paragraph{Supplementary Figure 2} \label{spf:imputation}
Average relative error (in absolute value) for the missing data imputation
algorithms with individual missing values amounting to 2\%, 5\%, 10\% and 20\%
of the total.

\paragraph{Supplementary Figure 3} \label{spf:imputation2}
Average relative error (in absolute value) for the missing data imputation
algorithms with values missing in 1-month batches (4 consecutive weeks)
amounting to 2\%, 5\%, 10\% and 20\% of the total.

\paragraph{Supplementary Figure 4} \label{spf:variance}
Proportion of the variance of each condition explained by the US states, by the
counties and by the counties together with the temporal autocorrelation. The
average for each of them over all conditions is reported at the bottom.

\end{document}